\documentclass[aps,prd,preprint,preprintnumbers,superscriptaddress,showpacs,tighten,nofootinbib]{revtex4}
\usepackage{amssymb}
\usepackage{mathrsfs}
\usepackage{mathbbold}
\usepackage{graphicx}
\usepackage{amsmath}
\usepackage{epsfig}
\usepackage{mciteplus}
\usepackage{amsmath,color}
\def\slc#1{\setbox0=\hbox{$#1$}           
    \dimen0=\wd0                                 
    \setbox1=\hbox{/} \dimen1=\wd1               
    \ifdim\dimen0>\dimen1                        
       \rlap{\hbox to \dimen0{\hfil/\hfil}}      
       #1                                        
    \else                                        
       \rlap{\hbox to \dimen1{\hfil$#1$\hfil}}   
       /                                         
    \fi}

\begin{document}

\title{Radiative corrections and explicit perturbations to the
tetra-maximal neutrino mixing with large $\theta_{13}$}

\author{He Zhang}
\affiliation{Max-Planck-Institut f\"{u}r Kernphysik, Postfach
103980, D-69029 Heidelberg, Germany} \email{he.zhang@mpi-hd.mpg.de}

\author{Shun Zhou}
\affiliation{Max-Planck-Institut f{\"u}r Physik
(Werner-Heisenberg-Institut), F{\"o}hringer Ring 6, D-80805
M{\"u}nchen, Germany} \email{zhoush@mppmu.mpg.de}


\preprint{MPP-2011-79}

\begin{abstract}
\noindent The tetra-maximal neutrino mixing pattern predicts a
relatively large reactor mixing angle $\theta_{13} \approx
8.4^\circ$, which is in good agreement with the latest best-fit
value $\theta_{13} = 9^\circ$. However, its prediction for
$\theta_{12} \approx 30.4^\circ$ is inconsistent with current
oscillation data at the $3\sigma$ C.L. We show that explicit
perturbations to the tetra-maximal mixing can naturally enhance
$\theta_{12}$ to its best-fit value $\theta_{12} = 34^\circ$.
Furthermore, we demonstrate that if the tetra-maximal mixing is
produced by a certain flavor symmetry at a high-energy scale
$\Lambda = 10^{14}~{\rm GeV}$, significant radiative corrections in
the minimal supersymmetric standard model can modify $\theta_{12}$
to be compatible with experimental data at the electroweak scale
$\Lambda_{\rm EW} = 10^2~{\rm GeV}$. The predictions for
$\theta_{13} \approx 8.4^\circ$ and $\theta_{23} = 45^\circ$, as
well as the CP-violating phases $\rho = \sigma = -\delta =
90^\circ$, are rather stable against radiative corrections.
\end{abstract}


\maketitle

\section{Introduction}

Recent solar, atmospheric, reactor and accelerator neutrino
experiments have provided us with compelling evidence that neutrinos
are massive and lepton flavors are mixed~\cite{Strumia:2006db}. The
lepton flavor mixing is described by a $3\times 3$ unitary matrix
$V$, the Maki-Nakagawa-Sakata (MNS) matrix~\cite{Maki:1962mu}, which
can be parametrized by three rotation angles and three CP-violating
phases. In the standard parametrization advocated by the Particle
Data Group~\cite{Nakamura:2010zzi} and in
Ref.~\cite{Fritzsch:2001ty,*Xing:2003ez}, the MNS matrix reads
\begin{equation}
V = \left(\begin{matrix}c_{12} c_{13} & s_{12} c_{13} & s_{13}
e^{-i\delta} \cr -s_{12} c_{23} - c_{12} s_{23} s_{13}e^{i\delta} &
c_{12} c_{23} - s_{12} s_{23} s_{13}e^{i\delta} & s_{23} c_{13} \cr
s_{12} s_{23} - c_{12} c_{23} s_{13}e^{i\delta} & -c_{12} s_{23} -
s_{12} c_{23} s_{13}e^{i\delta} & c_{23} c_{13}
\end{matrix}\right) \left(\begin{matrix}e^{i\rho} & 0 & 0 \cr 0 &
e^{i\sigma} & 0 \cr 0 & 0 & 1\end{matrix}\right) \; ,
\end{equation}
where $s_{ij} \equiv \sin \theta_{ij}$ and $c_{ij} \equiv \cos
\theta_{ij}$ (for $ij = 12, 23, 13$). The latest global analysis of
current neutrino oscillation data yields $31^\circ < \theta_{12} <
37^\circ$, $36^\circ < \theta_{23} < 53^\circ$ and $4^\circ <
\theta_{13} < 13^\circ$ at the $3\sigma$ C.L., and the best-fit
values of three mixing angles $\theta_{12} = 34^\circ$, $\theta_{23}
= 40^\circ$ and $\theta_{13} = 9^\circ$~\cite{Fogli:2011qn}. Note
that the global analysis has shown more than $3\sigma$ evidence for
a non-vanishing reactor mixing angle $\theta_{13} \neq 0$, while the
maximal atmospheric mixing $\theta_{23} = 45^\circ$ is still allowed
at the $1\sigma$ C.L. However, three CP-violating phases $(\delta,
\rho, \sigma)$ are entirely unconstrained. The smallest mixing angle
$\theta_{13}$ and the Dirac CP-violating phase $\delta$ will be
measured in the ongoing and forthcoming neutrino oscillation
experiments, while the Majorana CP-violating phases $\rho$ and
$\sigma$ can be constrained in the neutrinoless double beta decay
experiments and colliders.

How to understand the lepton flavor mixing pattern remains an open
question in elementary particle physics. Based on the observed
neutrino mixing angles, however, several interesting constant mixing
patterns have been proposed and widely discussed in the context of
flavor symmetries. For instance, the tri-bimaximal mixing pattern
with $\theta_{12} \approx 35.3^\circ$, $\theta_{23} = 45^\circ$ and
$\theta_{13} = 0$ is in good agreement with current oscillation
data~\cite{Harrison:2002er,*Xing:2002sw,*Harrison:2002kp,*He:2003rm}.
Its predictions of $\theta_{23} = 45^\circ$ and $\theta_{13} = 0$
have motivated a torrent of activities in the model building with
discrete flavor symmetries, which give rise to the tri-bimaximal
neutrino mixing at the leading order (see, e.g.,
\cite{Ishimori:2010au,*Albright:2010ap} and references therein).
Nevertheless, the latest result from the T2K
experiment~\cite{Abe:2011sj}, in which $\nu_\mu \to \nu_e$
oscillations have been observed, indicates $5.0^\circ < \theta_{13}
< 16.0^\circ$ for the normal mass hierarchy and $5.8^\circ <
\theta_{13} < 17.8^\circ$ for the inverted mass hierarchy at the
$90\%$ C.L. The best-fit value from T2K data is $\theta_{13} \approx
10^\circ$, which is also consistent with the global-fit analysis.
Moreover, the MINOS experiment has recently reported the observation
of $\nu_\mu \to \nu_e$ oscillations, disfavoring the assumption of
$\theta_{13} = 0$ at the $1.5\sigma$ C.L. Possible ways to realize a
relatively large $\theta_{13}$ have recently been discussed in
Ref.~\cite{Chang:2009mv,*Xing:2010pn,*He:2011kn,*Shimizu:2011xg,*Xing:2011at,*He:2011gb,*Zhou:2011nu,*Araki:2011wn,*Haba:2011nv,*Meloni:2011fx,*Morisi:2011wn,*Chao:2011sp}.
Considering the experimental tendency of a non-vanishing or even
relatively-large $\theta_{13}$, we argue that the tetra-maximal
mixing pattern~\cite{Xing:2008ie}
\begin{equation}
\widehat{V} = \frac{1}{2} \left(\begin{matrix} \displaystyle 1 +
\frac{1}{\sqrt{2}} & 1 & \displaystyle 1 - \frac{1}{\sqrt{2}} \cr
\displaystyle -\frac{1}{\sqrt{2}}\left[1 + i(1 -
\frac{1}{\sqrt{2}})\right] & \displaystyle 1 + i\frac{1}{\sqrt{2}} &
\displaystyle \frac{1}{\sqrt{2}}\left[1 - i(1 +
\frac{1}{\sqrt{2}})\right] \cr \displaystyle
-\frac{1}{\sqrt{2}}\left[1 - i(1 - \frac{1}{\sqrt{2}})\right] &
\displaystyle 1 - i\frac{1}{\sqrt{2}} & \displaystyle
\frac{1}{\sqrt{2}}\left[1 + i(1 + \frac{1}{\sqrt{2}})\right]
\end{matrix}\right) \; ,
\end{equation}
with $\theta_{12} \approx 30.4^\circ$, $\theta_{23} = 45^\circ$, and
$\theta_{13} \approx 8.4^\circ$ may serve as a better starting point
to search for the true symmetry underlying the lepton flavor mixing.
The prediction of $\theta_{13} \approx 8.4^\circ$ from the
tetra-maximal mixing pattern is in excellent agreement with the
latest neutrino oscillation data, while that of $\theta_{23} =
45^\circ$ and the maximal CP-violating phase $\delta = - 90^\circ$
may hint at a certain flavor symmetry.

In the present work, we consider possible deviations from the
tetra-maximal mixing pattern. The motivation for such an
investigation is two-fold. First, the solar mixing angle predicted
by the tetra-maximal mixing pattern is $\theta_{12} \approx
30.4^\circ$, which is much smaller than the best-fit value
$\theta_{12} = 34^\circ$ and even not lying in the $3\sigma$ range
from the global analysis. Second, if the tetra-maximal mixing
pattern is obtained by assuming a certain flavor symmetry, which in
general works at a high-energy scale, the mixing angles will receive
significant radiative corrections when running from the symmetry
scale to the low-energy scale, at which the mixing angles are
actually measured in oscillation experiments. One immediate question
is whether it is possible to increase $\theta_{12}$ to the observed
value by introducing explicit perturbations to the tetra-maximal
mixing pattern or by taking into account the radiative corrections,
while both $\theta_{23}$ and $\theta_{13}$ remain consistent with
experimental data. We have found the answer is affirmative.

The remaining part of this work is organized as follows. In Sec.~II,
we show that only one perturbation parameter is enough to increase
$\theta_{12}$ to its best-fit value, and leads to an even larger
$\theta_{13}$. Furthermore, we demonstrate that both the maximal
atmospheric mixing angle $\theta_{23} = 45^\circ$ and the maximal
CP-violating phase $\delta = -90^\circ$ are not affected by the
perturbation to any order. In Sec.~III, we explicitly solve the
renormalization group equations (RGE's) for neutrino mixing
parameters within the minimal supersymmetric standard model (MSSM),
and confirm that the observed $\theta_{12}$ can be obtained at the
electroweak scale $\Lambda_{\rm EW} = 100~{\rm GeV}$ from the
tetra-maximal mixing pattern at the high-energy scale $\Lambda =
10^{14}~{\rm GeV}$. On the other hand, both $\theta_{13}$ and
$\theta_{23}$, as well as three CP-violating phases $(\delta, \rho,
\sigma)$, are rather stable against the radiative corrections.
Finally, we summarize our conclusions in Sec.~IV.

\section{Explicit Perturbations}

First of all, we briefly recall the tetra-maximal mixing pattern and
its salient features. As shown in Ref.~\cite{Xing:2008ie}, the
tetra-maximal mixing matrix in Eq.~(2) can be decomposed into four
maximal rotations
\begin{equation}
\widehat{V} = P_l \otimes O_{23}(\pi/4, \pi/2) \otimes
O_{13}(\pi/4,0) \otimes O_{12}(\pi/4,0) \otimes O_{13}(\pi/4,\pi) \;
,
\end{equation}
where $P_l = {\rm Diag}\{1, 1, i\}$, and
$O_{ij}(\theta_{ij},\delta_{ij})$ is a rotation with the angle
$\theta_{ij}$ and the phase $\delta_{ij}$ in the complex $i$-$j$
plane for $ij = 12, 23, 13$. Comparing between Eq.~(1) and Eq.~(2),
one can extract three neutrino mixing angles
\begin{eqnarray}
\tan \theta_{12} = 2 - \sqrt{2} \; , ~~~ \tan \theta_{23} = 1 \; ,
~~~ \sin \theta_{13} = \frac{1}{4}(2 - \sqrt{2}) \; ,
\end{eqnarray}
or explicitly $\theta_{12} \approx 30.4^\circ$, $\theta_{23} =
45^\circ$, $\theta_{13} \approx 8.4^\circ$. Furthermore, three
CP-violating phases are $\rho = \sigma = -\delta = 90^\circ$ can be
obtained by redefining the phases of charged-lepton fields and
recasting the tetra-maximal mixing pattern into the standard form in
Eq.~(1).

Although $\theta_{13} \approx 8.4^\circ$ is relatively large and
well compatible with the recent T2K and MINOS results, $\theta_{12}
\approx 30.4^\circ$ is much smaller than the best-fit value
$\theta_{12} \approx 34^\circ$, and even not covered in the
$3\sigma$ range, i.e., $31^\circ < \theta_{12} < 37^\circ$. In order
to increase $\theta_{12}$ significantly but not change much both
$\theta_{23}$ and $\theta_{13}$ such that all three mixing angles
become consistent with current oscillation data, we have to slightly
modify the tetra-maximal mixing pattern in a proper manner. One
straightforward way is just to introduce a small perturbation to the
maximal rotation in the $1$-$2$ complex plane. In this case, the MNS
matrix turns out to be
\begin{equation}
V = P_l \otimes O_{23}(\pi/4, \pi/2) \otimes O_{13}(\pi/4,0) \otimes
O_{12}(\pi/4 + \varepsilon_{12},0) \otimes O_{13}(\pi/4,\pi) \; ,
\end{equation}
with $\varepsilon_{12} \ll 1$. In order to see how the neutrino
mixing angles are changed, we expand the MNS matrix in Eq.~(5) with
respect to the perturbation parameter to the second order
\begin{eqnarray}
V = \widehat{V} &+& \frac{1}{4} \varepsilon_{12}
\left(\begin{matrix}-\sqrt{2} & 2 & \sqrt{2} \cr -\sqrt{2} - i & -2
+ i\sqrt{2} & \sqrt{2} + i \cr -\sqrt{2} + i & -2
- i\sqrt{2} & \sqrt{2} - i \end{matrix}\right) \nonumber \\
&+& \frac{1}{8} \varepsilon^2_{12} \left(\begin{matrix}-\sqrt{2} &
-2 & \sqrt{2} \cr \sqrt{2} - i & -2 - i\sqrt{2} & -\sqrt{2} + i \cr
\sqrt{2} + i & -2 + i\sqrt{2} & -\sqrt{2} - i \end{matrix}\right) +
{\cal O}(\varepsilon^3_{12})\; .
\end{eqnarray}
Some discussions are in order:
\begin{enumerate}
\item Comparing Eq.~(6) with the standard parametrization in Eq.~(1), one can
immediately derive three neutrino mixing angles
\begin{eqnarray}
\tan \theta_{12} &=& (2-\sqrt{2}) \left[1 + \sqrt{2}
\varepsilon_{12} - \frac{3\sqrt{2} - 4}{2} \varepsilon^2_{12}
\right] + {\cal O}(\varepsilon^3_{12}) \; , \nonumber \\
\sin\theta_{13} &=& \frac{2-\sqrt{2}}{4} \left[1 + (\sqrt{2}+1)
\varepsilon_{12} + \frac{\sqrt{2}+1}{2}\varepsilon^2_{12}\right] +
{\cal O}(\varepsilon^3_{12}) \; ,
\end{eqnarray}
and $\tan \theta_{23} = 1$. Taking the best-fit value of
$\theta_{12} = 34^\circ$, we find $\varepsilon_{12} \approx 0.11$
from the first identity in Eq.~(7). Inserting $\varepsilon_{12}
\approx 0.11$ into the second identity, one can obtain $\theta_{13}
\approx 10.8^\circ$, which is in good agreement with the T2K
result~\cite{Abe:2011sj} and close to the $3\sigma$ upper limit
$\theta_{13} < 13^\circ$ from the global-fit
analysis~\cite{Fogli:2011qn}.

\item Note that the maximal atmospheric mixing angle, i.e., $\theta_{23} =
45^\circ$, is not modified, because there is a generalized
permutation symmetry in the tetra-maximal mixing matrix in Eq.~(3)
and its perturbed version in Eq.~(5): $P_{23} U = U^*$ for both $U =
\widehat{V}$ and $U = V$, where $P_{23}$ denotes the exchange of the
second and third rows. Such a permutation symmetry originates from
the pure phase matrix $P_l$ and the first maximal rotation
$O_{23}(\pi/4, \pi/2)$, which are the only complex matrices. Hence
we have $\tan \theta_{23} = |V_{\mu 3}|/|V_{\tau 3}| = 1$ that is
independent of the perturbation.

\item The tetra-maximal mixing pattern predicts the maximal CP-violating
phase $\delta = -90^\circ$ and the Jarlskog invariant ${\cal J}
\equiv {\rm Im} \left[V_{e2} V_{\mu 3} V^*_{e3} V^*_{\mu 2}\right] =
-1/32$ for leptonic CP violation~\cite{Xing:2008ie}. Taking account
of the perturbation, we arrive at
\begin{equation}
{\cal J} = -\frac{1}{32} \left[1 + 3\varepsilon_{12} + \frac{3}{2}
\varepsilon^2_{12}\right] + {\cal O}(\varepsilon^3_{12}) \; .
\end{equation}
Therefore, the magnitude of leptonic CP violation, measured by
$|{\cal J}|$, is enhanced due to a positive $\varepsilon_{12}$,
which has been implemented to increase both $\theta_{12}$ and
$\theta_{13}$. However, we can prove that the maximal CP-violating
phase $\delta = -90^\circ$ is maintained to any order of
perturbations. The proof is as follows:
\begin{enumerate}
\item Since the matrix elements $V_{e i}$ (for $i = 1, 2, 3$)
are always real, the Jarlskog invariant is ${\cal J} \equiv {\rm Im}
\left[V_{e2} V_{\mu 3} V^*_{e3} V^*_{\mu 2}\right] = V_{e 2} V_{e 3}
{\rm Im} \left[V^*_{\mu 2} V_{\mu 3}\right]$. On the other hand, we
have ${\cal J} = s_{12} c_{12} s_{23} c_{23} s_{13} c^2_{13} \sin
\delta$ in the standard parametrization, or equivalently ${\cal J} =
|V_{e1}| |V_{e2}| |V_{e3}| \sin \delta/2$, where $\tan \theta_{23} =
1$ has been input. Therefore, the CP-violating phase is determined
by
\begin{equation}
\sin \delta = \frac{2 V_{e 2} V_{e 3} {\rm Im}\left[V^*_{\mu 2}
V_{\mu 3} \right]}{|V_{e1}| |V_{e2}| |V_{e3}|}  \; .
\end{equation}

\item The unitarity of the MNS matrix $V$ leads to the normalization
condition $|V_{\mu 1}|^2 + |V_{\mu 2}|^2 + |V_{\mu 3}|^2= 1$ and the
orthogonality condition $V^2_{\mu 1} + V^2_{\mu 2} + V^2_{\mu 3} =
0$. The latter condition is guaranteed by $P_{23} V = V^*$. Thus we
obtain
\begin{equation}
\left|V_{\mu 1}\right|^4 = \left|V_{\mu 2}\right|^4 + \left|V_{\mu
3}\right|^4 + 2\left[\left({\rm Re}\left[V^*_{\mu 2} V_{\mu 3}
\right]\right)^2 - \left({\rm Im}\left[V^*_{\mu 2} V_{\mu 3}
\right]\right)^2\right] \; ,
\end{equation}
and
\begin{equation}
\left(1 - \left|V_{\mu 1}\right|^2\right)^2 = \left|V_{\mu
2}\right|^4 + \left|V_{\mu 3}\right|^4 + 2\left[\left({\rm
Re}\left[V^*_{\mu 2} V_{\mu 3} \right]\right)^2 + \left({\rm
Im}\left[V^*_{\mu 2} V_{\mu 3} \right]\right)^2\right] \; .
\end{equation}
Subtracting Eq.~(10) from Eq.~(11) and using another normalization
relation $|V_{e1}|^2 + 2|V_{\mu 1}|^2 = 1$, one can verify $|V_{e1}|
= 2\left|{\rm Im}\left[V^*_{\mu 2} V_{\mu 3} \right]\right|$, which
together with Eq.~(9) leads to $|\sin \delta| = 1$. As long as the
perturbations are small, the sign of ${\cal J}$ is determined by the
tetra-maximal mixing matrix $\widehat{V}$ and thus $\delta =
-90^\circ$ is valid to any order of perturbations.
\end{enumerate}

\item In the basis where the charged-lepton mass matrix is diagonal,
we can reconstruct the neutrino mass matrix by the MNS matrix $V$
and neutrino masses $m_i$ (for $i = 1, 2, 3$)
\begin{equation}
M_{\nu} = V \left(\begin{matrix}m_1 & 0 & 0 \cr 0 & m_2 & 0 \cr 0 &
0 & m_3 \end{matrix}\right) V^{\rm T} \; .
\end{equation}
With the help of the identity $P_{23} V = V^*$, it is
straightforward to show that
\begin{equation}
M^*_\nu = V^* \widehat{m} V^\dagger = (P_{23} V) \widehat{m} (P_{23}
V)^{\rm T} = P_{23} M_\nu P^{\rm T}_{23} \; ,
\end{equation}
where $P^{\rm T}_{23} = P^{-1}_{23} = P_{23}$ and $\widehat{m} =
{\rm Diag}\{m_1, m_2, m_3\}$. As a consequence of Eq.~(13), we can
get
\begin{equation}
M_{ee} = M^*_{ee} \; , ~~~ M_{e\tau} = M^*_{e \mu} \; , ~~~ M_{\tau
\tau} = M^*_{\mu \mu} \; , ~~~ M_{\mu \tau} = M^*_{\mu \tau} \; ,
\end{equation}
where $M_{\alpha \beta}$ denotes the matrix element of $M_\nu$ (for
$\alpha, \beta = e, \mu, \tau$). Such a special structure of
$M_\nu$, which can give rise to both maximal atmospheric mixing
$\theta_{23} = 45^\circ$ and maximal CP-violating phase $\delta =
\pm 90^\circ$, may result from a certain flavor symmetry. Since both
$V$ and the tetra-maximal mixing matrix $\widehat{V}$ share the same
permutation symmetry, the perturbation under consideration
does not spoil the symmetry of neutrino mass matrix in
Eq.~(14).
\end{enumerate}

If we generalize the perturbation scheme in Eq.~(5) and add small
perturbations to all four maximal rotations in Eq.~(3), the MNS
matrix can be written as
\begin{eqnarray}
V =\widehat{V} &+& \frac{1}{4} \varepsilon_{12}
\left(\begin{matrix}-\sqrt{2} & 2 & \sqrt{2} \cr -\sqrt{2} - i & -2
+ i\sqrt{2} & \sqrt{2} + i \cr -\sqrt{2} + i & -2
- i\sqrt{2} & \sqrt{2} - i \end{matrix}\right) \nonumber \\
&+& \frac{1}{4} \varepsilon_{13} \left(\begin{matrix} 4 - 2\sqrt{2}
& -2 & 0 \cr \sqrt{2} & + i\sqrt{2} & \sqrt{2} + 2i(\sqrt{2} - 1)
\cr\sqrt{2} & - i \sqrt{2}& \sqrt{2} - 2i(\sqrt{2} - 1)
\end{matrix}\right)
\nonumber \\
&+& \frac{1}{4} \varepsilon_{23} \left(\begin{matrix}0 & 0 & 0 \cr +
\sqrt{2} - i(\sqrt{2} - 1) & -2 + i\sqrt{2} & -\sqrt{2} - i
(\sqrt{2} + 1)\cr -\sqrt{2} -i(\sqrt{2} - 1)  & + 2  + i\sqrt{2}  &
+ \sqrt{2} - i(\sqrt{2} + 1) \end{matrix}\right) \; ,
\end{eqnarray}
where the higher-order terms ${\cal O}(\varepsilon^2)$ have been
neglected. In this case, the neutrino mixing angles receive
corrections from all these perturbations. Comparing Eq.~(15) with
the standard parametrization in Eq.~(1), we obtain
\begin{eqnarray}
\tan \theta_{23} &=& 1 + \frac{2}{17} \left(8\sqrt{2} - 3\right)
\varepsilon_{23} \; , \nonumber \\
\sin \theta_{13} &=& \frac{2-\sqrt{2}}{4}\left[1 +
(\sqrt{2}+1)\varepsilon_{12} \right] \; , \nonumber \\
\tan \theta_{12} &=& (2-\sqrt{2}) \left[1 + \sqrt{2}
\varepsilon_{12} - (7 - 4\sqrt{2}) \varepsilon_{13} \right] \;,
\end{eqnarray}
to the first order of perturbations. Since $\varepsilon_{13}$
contributes only to the solar mixing angle $\theta_{12}$, we can
switch off both $\varepsilon_{12}$ and $\varepsilon_{23}$, and
choose $\varepsilon_{13} \approx -0.11$ to obtain the best-fit value
of $\theta_{12} = 34^\circ$. The unique and efficient way to enhance
both $\theta_{12}$ and $\theta_{13}$ is to adjust the perturbation
parameter $\varepsilon_{12}$ as we have done before. Now that
$\varepsilon_{23}$ breaks the permutation symmetry $P_{23} V = V^*$,
it induces deviation of $\theta_{23}$ from the maximal mixing, and
also invalidates the maximal CP-violating phase. In addition, the
symmetry relations for the reconstructed neutrino mass matrix in
Eq.~(14) are not respected.

\section{Radiative Corrections}

Now we proceed to consider another possible deviation from the
tetra-maximal mixing pattern, i.e., the renormalization group (RG)
corrections. As already mentioned in Sec.~I, the flavor symmetries
generating the tetra-maximal mixing are generally preserved at
high-energy scales, such as the grand unification scale (e.g.,
$\Lambda_{\rm GUT} = 10^{16}~{\rm GeV}$) or the hypothetical seesaw
scale (e.g., $\Lambda = 10^{14}~{\rm GeV}$), while the neutrino
mixing parameters are determined or constrained in neutrino
oscillation experiments at low-energy scales. The gap between the
high-energy predictions and the low-energy measurements is bridged
by the RG evolution. The RG running effects may then serve as an
explanation for the discrepancy between the flavor symmetric mixing
pattern and observables.

The RGE's for neutrino mixing parameters have been derived within
various theoretical
frameworks~\cite{Chankowski:1993tx,*Babu:1993qv,*Antusch:2001ck,
*Antusch:2001vn,*Chao:2006ye,*Schmidt:2007nq,*Chakrabortty:2008zh,
*Blennow:2011mp}.
In the supersymmetric theories with a large $\tan\beta$, it has been
found that the RG evolution may lead to significant modifications to
the mixing parameters, in particular the solar mixing angle
$\theta_{12}$~(see e.g., Ref.~\cite{Ray:2010rz} and references
therein). To be explicit, we write down the RGE's for neutrino
mixing angles in the MSSM in the leading-order
approximation~\cite{Antusch:2003kp},
\begin{eqnarray}
\Dot{\theta}_{12} & \approx & -\frac{y_\tau^2 s^2_{12} c^2_{12}
s_{23}^2 }{8\pi^2 \Delta m^2_{\rm sol}} \left[m^2_1 + m^2_2 + 2 m_1
m_2 c_{2(\rho - \sigma)}\right] \;,
\nonumber \\
\Dot{\theta}_{13} & \approx & + \frac{y_\tau^2 s^2_{12}c^2_{12}
s^2_{23} c^2_{23} m_3}{2\pi^2\Delta m^2_{\rm
atm}\left(1+\zeta\right)} \left[ m_1 c_{(2\rho+\delta)} -
\left(1+\zeta\right) m_2 c_{(2\sigma+\delta)} - \zeta m_3 c_\delta
\right] \;,
\nonumber \\
\Dot{\theta}_{23} & \approx & -\frac{y_\tau^2 s^2_{23} c^2_{23}
}{8\pi^2\Delta m^2_{\rm atm}} \left[c_{12}^2 \left(m^2_2 + m^2_3 +
2m_2 m_3 c_{2\sigma}\right) + \frac{s_{12}^2 \left(m^2_1 + m^2_3 + 2
m_1 m_3 c_{2\rho}\right)}{1+\zeta} \right] \;,
\end{eqnarray}
where $\Dot{\theta}_{ij} = {\rm d}\theta_{ij}/{\rm d}t$ with $t =
\ln (\mu/\mu_0)$, $\zeta \equiv \Delta m^2_{\rm sol}/\Delta m^2_{\rm
atm}$ with $\Delta m^2_{\rm sol} \equiv m^2_2 - m^2_1 \approx
7.6\times 10^{-5}~{\rm eV}^2$ and $|\Delta m^2_{\rm atm}| \equiv
|m^2_3 - m^2_2| \approx 2.3 \times 10^{-3}~{\rm eV}^2$
\cite{Fogli:2011qn}, and $y_\tau$ denotes the Yukawa coupling of tau
charged lepton. Note that the terms of ${\cal O}(\theta_{13})$ have
been safely neglected in Eq.~(17). In addition, we have defined
$c_{2(\rho - \sigma)} \equiv \cos 2(\rho - \sigma)$, $c_{2\rho}
\equiv \cos 2\rho$ and so on. Some general features of the RGE's are
summarized as follows:
\begin{itemize}
\item Since $|\Delta m^2_{\rm atm}|/\Delta m^2_{\rm sol} \approx 30$, one
can observe that $\theta_{12}$ in general receives more remarkable
RG corrections than $\theta_{23}$ and $\theta_{13}$. Furthermore,
when running from a high-energy scale $\Lambda = 10^{14}~{\rm GeV}$
to the electroweak scale $\Lambda_{\rm EW} = 100~{\rm GeV}$, the
radiative corrections to $\theta_{12}$ are always positive, i.e.,
$\theta_{12}(\Lambda) < \theta_{12}(\Lambda_{\rm EW})$, which is
independent of the neutrino mass hierarchy and the CP-violating
phases. Therefore, if the tri-bimaximal mixing pattern with
$\theta_{12} \approx 35.3^\circ$ is assumed at the high-energy
scale, the RG effects will lead to an even larger $\theta_{12}$ at
the low-energy scale, which may be in conflict with current
oscillation data~\cite{Luo:2005fc}. Note that $\theta_{12}$ receives
negative corrections in the standard model, but there are no visible
RG effects in this case due to the absence of $\tan \beta$
enhancement.

\item As for $\theta_{23}$ and $\theta_{13}$, the RG corrections could
be either positive or negative, depending on the neutrino mass
hierarchies. More explicitly, we have $\Dot{\theta}_{23} < 0$ in the
normal hierarchy (NH) case with $m_3 > m_2 > m_1$, while
$\Dot{\theta}_{23} > 0$ in the inverted hierarchy (IH) case with
$m_2 > m_1 > m_3$. The sign of $\Dot{\theta}_{13}$ further depends
on the CP-violating phases.
\end{itemize}

On the other hand, the RGE's for the CP-violating phases are
approximately given by~\cite{Antusch:2003kp}
\begin{eqnarray}
\Dot{\delta} &=& \frac{y_\tau^2 s^2_{12} c^2_{12} s^2_{23} c^2_{23}
m_3 \theta^{-1}_{13}}{2\pi^2 \Delta m^2_\mathrm{atm}
\left(1+\zeta\right)} \left[\left(1 + \zeta\right) m_2
s_{(2\sigma+\delta)} - m_1 s_{(2\rho+\delta)} + \zeta m_3 s_\delta
\right] \;,
\nonumber \\
\Dot{\rho} &=& \frac{y_\tau^2}{8\pi^2} \left\{m_3
(c^2_{23}-s^2_{23}) \frac{m_1 s_{12}^2 s_{2\rho} + \left( 1+\zeta
\right) m_2 c_{12}^2 s_{2\sigma}}{\Delta m^2_\mathrm{atm} \left(
1+\zeta \right)} + \frac{m_1 m_2 c_{12}^2 s_{23}^2
s_{2(\rho-\sigma)}}{\Delta m^2_\mathrm{sol}} \right\} \;,
\nonumber \\
\Dot{\sigma} &=& \frac{y_\tau^2}{8\pi^2} \left\{m_3
(c^2_{23}-s^2_{23}) \frac{m_1 s_{12}^2 s_{2\rho} + \left( 1+\zeta
\right) m_2 c_{12}^2 s_{2\sigma}}{\Delta m^2_\mathrm{atm} \left(
1+\zeta \right)} + \frac{m_1 m_2 s_{12}^2 s_{23}^2
s_{2(\rho-\sigma)}}{\Delta m^2_\mathrm{sol}} \right\} \; ,
\end{eqnarray}
where the terms of ${\cal O}(\theta_{13})$ have been ignored. For
the tetra-maximal mixing pattern, one can insert the initial
condition $\rho = \sigma = -\delta = 90^\circ$ into Eq.~(18), and
immediately obtain $\Dot{\rho}  = \Dot{\sigma} = 0$ at the leading
order, indicating that all the Majorana phases are rather stable
against RG corrections. As for $\delta$, we have $\Dot{\delta}
\propto \frac{m_3(m_2-m_1)}{(m_2+m_3)(m_1+m_3)} \ll 1$ for any
neutrino mass hierarchies. Therefore, there is no enhancement factor
boosting the RG running of $\delta$, and $\delta=-90^\circ$ is also
stabilized at any energy scales. This has also been confirmed in our
numerical calculations.

\begin{figure}[t]
\vspace{-0.7cm}
\includegraphics[width=0.5\textwidth]{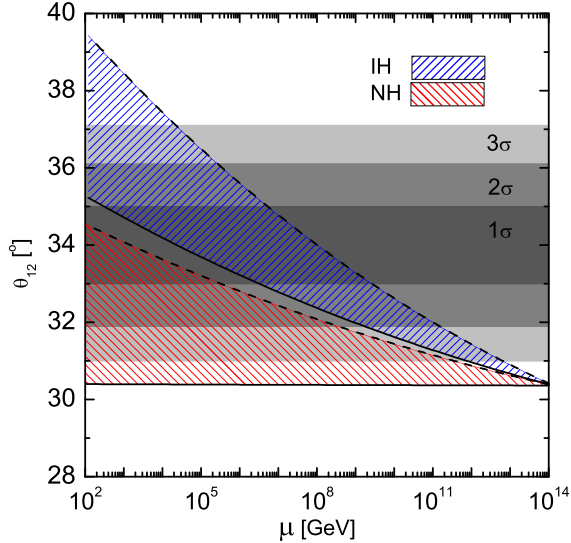}
\vspace{-0.4cm}
\caption{\label{fig:theta12} The RG evolution of $\theta_{12}$ in
the MSSM with $\tan\beta=15$. The solid and dashed lines correspond
to $m_1(\Lambda)=0$ and $m_1(\Lambda)=0.05~{\rm eV}$ in the NH case,
or $m_3(\Lambda)=0$ and $m_3(\Lambda)=0.05~{\rm eV}$ in the IH case,
where $\Lambda = 10^{14}~{\rm GeV}$ is a typical seesaw scale. The
colored bands indicate the RG corrections to the mixing angle if the
lightest neutrino mass is varying in the range of $(0 \sim
0.05)~{\rm eV}$. The allowed ranges of $\theta_{12}$ from the global
analysis~\cite{Fogli:2011qn} are also shown as shaded areas.}
\end{figure}
\begin{figure}[t]
\vspace{-0.7cm}
\includegraphics[width=0.49\textwidth]{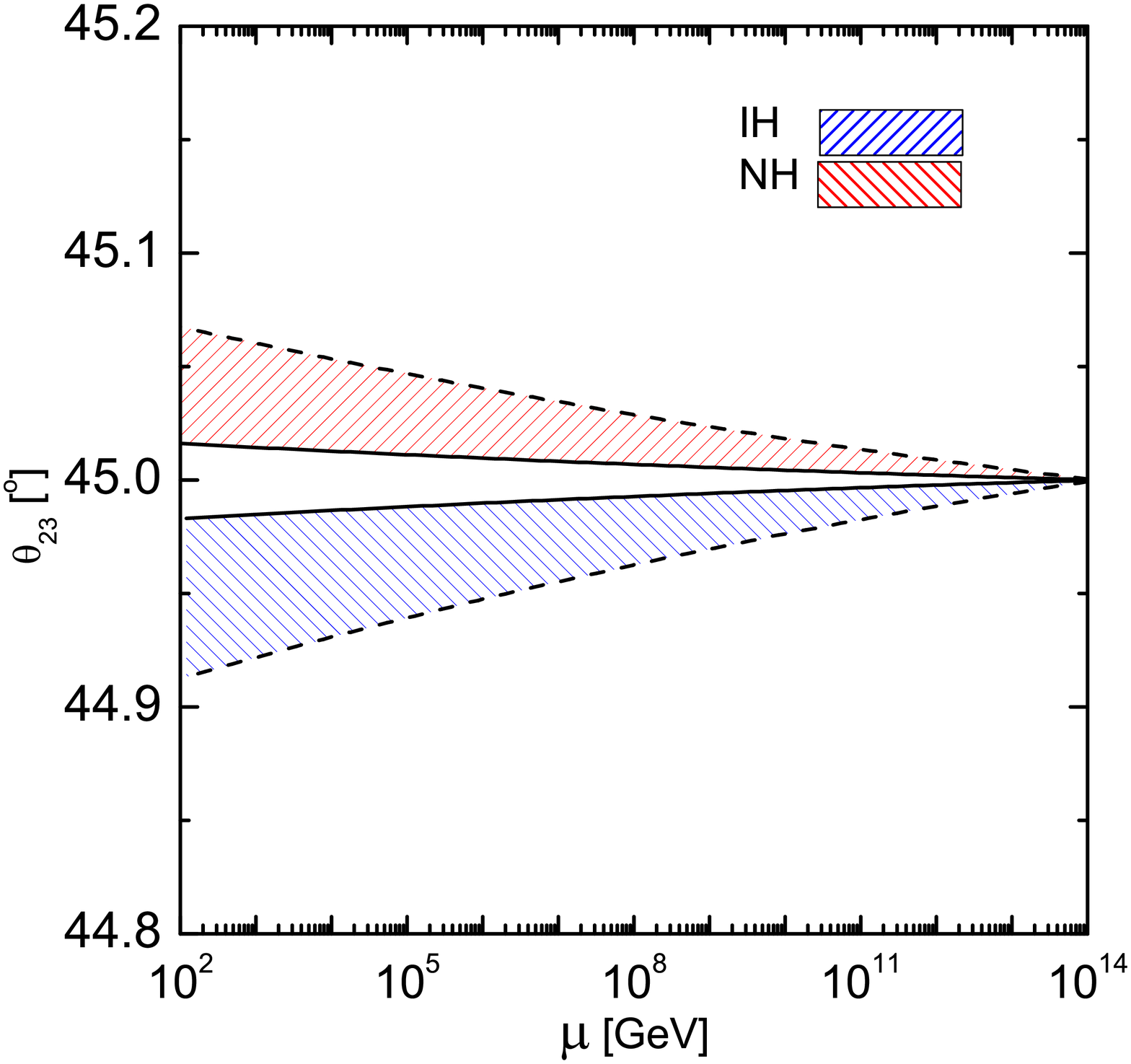}
\includegraphics[width=0.49\textwidth]{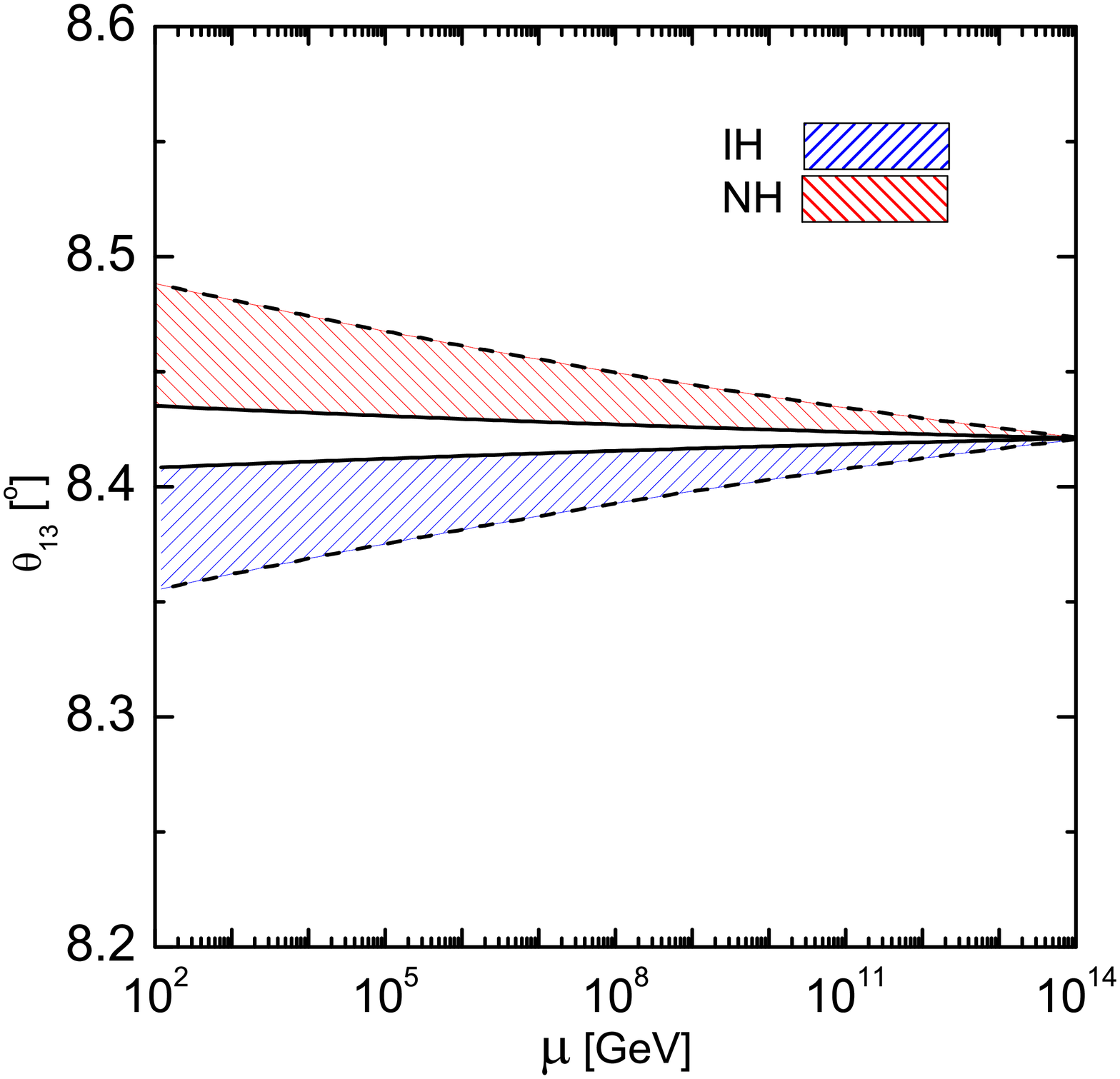}
\vspace{-0.4cm}
\caption{\label{fig:theta13} The RG evolution of $\theta_{13}$ and
$\theta_{23}$ in the MSSM with $\tan\beta=15$. The solid and dashed
lines correspond to $m_1(\Lambda)=0$ and $m_1(\Lambda)=0.05~{\rm
eV}$ in the NH case, or $m_3(\Lambda)=0$ and $m_3(\Lambda)=0.05~{\rm
eV}$ in the IH case, where $\Lambda = 10^{14}~{\rm GeV}$ is a
typical seesaw scale. The colored bands indicate the RG corrections
to the mixing angles if the lightest neutrino mass is varying in the
range of $(0 \sim 0.05)~{\rm eV}$.}
\end{figure}
\begin{figure}[t]
\vspace{-0.7cm}
\includegraphics[width=0.5\textwidth]{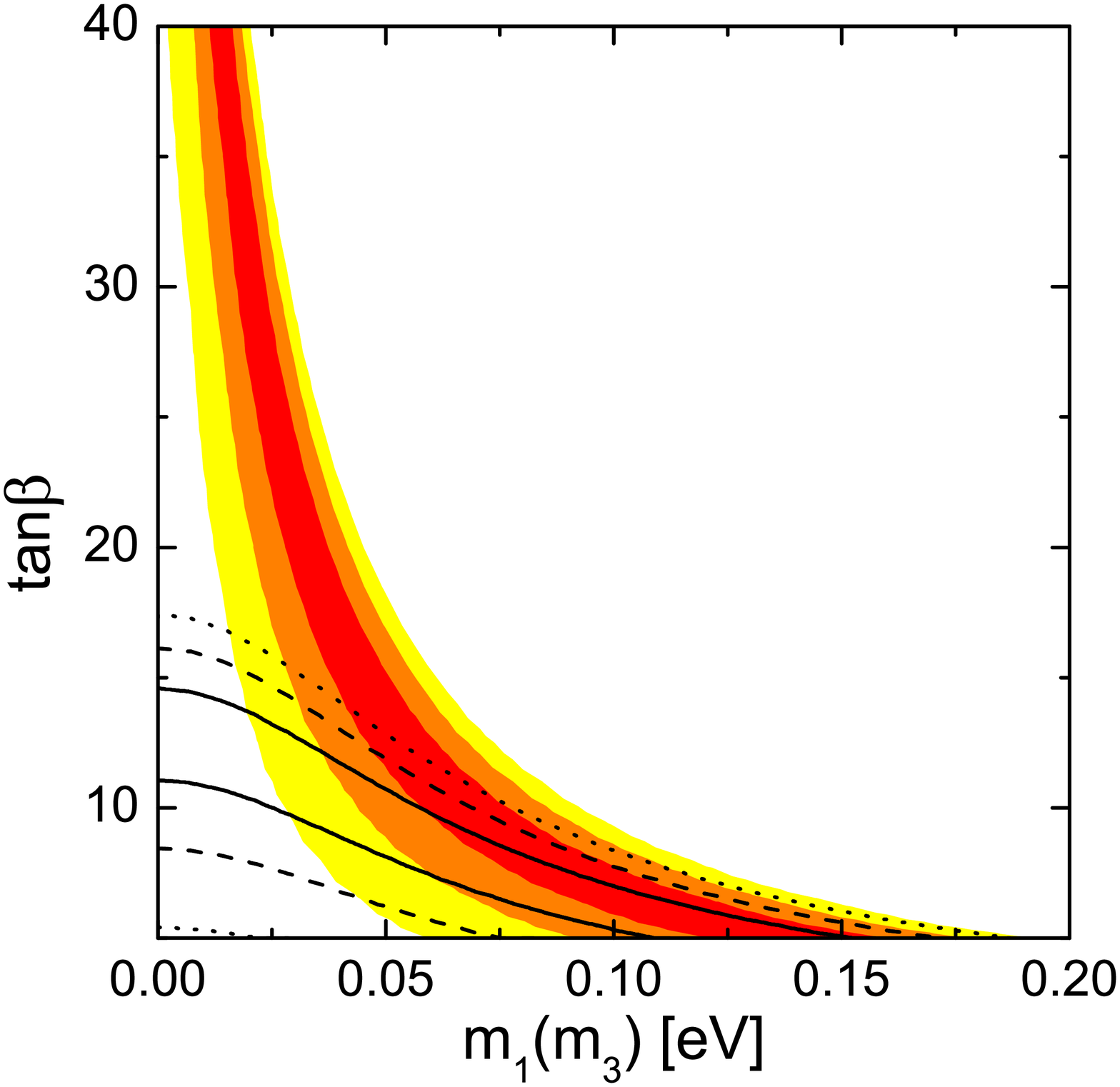}
\vspace{-0.4cm}
\caption{\label{fig:tanbeta} Allowed regions of $\tan\beta$ and
$m^{} _1$ in the NH case ($m_3$ in the IH case), where the
tetra-maximal mixing pattern is assumed at $\Lambda=10^{14}~{\rm
GeV}$ and the $\theta_{12}$ at $\Lambda_{\rm EW} = 10^2~{\rm GeV}$
is required to be in the 1$\sigma$, 2$\sigma$ and 3$\sigma$ ranges.
The shaded regions refer to the NH case, while the regions between
lines to the IH case.}
\end{figure}
To illustrate the RG corrections to the tetra-maximal mixing
pattern, we have numerically solved the full set of RGE's for
neutrino mixing angles. Some comments are in order:
\begin{enumerate}
\item In Fig.~1, the RG evolution of $\theta_{12}$ in the MSSM
with $\tan \beta = 15$ is shown. Note that we have assumed the
tetra-maximal mixing pattern at a cutoff scale $\Lambda
=10^{14}~{\rm GeV}$ (i.e., potentially the seesaw scale), and
allowed the lightest neutrino mass $m_1$ in the NH case (or $m_3$ in
the IH case) to vary in the range $(0 \sim 0.05)~{\rm eV}$. As seen
from Fig.~1, although the initial value of $\theta_{12}$ at the
cutoff scale deviates more than $3\sigma$ from its best-fit value,
the RG effects can enhance $\theta_{12}$ in a very efficient way so
as to fit the experimental data. In the NH case, $\theta_{12}$ can
be perfectly consistent with the low-scale measurements for the
chosen $\tan\beta$ and neutrino masses. In the IH case, the RG
corrections with $\tan\beta=15$ seem to be too large. This behavior
can be understood by noting that the first identity in Eq.~(17)
reduces to $\Dot{\theta}_{12} \propto (m_2+m_1)/(m_2 - m_1)$ in the
limit of $\rho = \sigma = 90^\circ$. In the NH case we have $(m_2 +
m_1)/(m_2 - m_1) \gtrsim 1$, whereas in the IH case $(m_2 +
m_1)/(m_2 - m_1) \gtrsim 10^2$ is expected. Therefore, a small
$\tan\beta$ (e.g., $\tan\beta < 20 $) is required in the IH case to
avoid the overlarge RG corrections.

\item In Fig.~2, we show the RG evolution of $\theta_{23}$ and
$\theta_{13}$ in the MSSM with $\tan \beta = 15$. Compared with the
case of $\theta_{12}$, the evolution of both $\theta_{23}$ and
$\theta_{13}$ is insignificant due to the suppression from $\Delta
m^2_{\rm sol}/|\Delta m^2_{\rm atm}|$. The radiative corrections to
$\theta_{23}$ and $\theta_{13}$ cannot exceed $0.1^\circ$ no matter
whether the NH or IH is assumed. It is worth mentioning that
$\Dot{\theta}_{13} = 0$ at the leading order, which can be seen by
inserting $\rho = \sigma = -\delta = 90^\circ$ into Eq.~(18). The
mild evolution of $\theta_{13}$ in Fig.~2 is actually attributed to
the high-order terms in the RGE's. Hence the predictions of
$\theta_{23} = 45^\circ$ and $\theta_{13} \approx 8.4^\circ$ from
the tetra-maximal mixing pattern are rather stable against radiative
corrections.

\item One may wonder if we can acquire significant RG corrections to
$\theta_{23}$ and $\theta_{13}$ by assuming a sizable $\tan\beta$
(i.e., $\tan\beta \gtrsim 20$) or a nearly degenerate mass spectrum
(i.e., $m_i \gtrsim 0.2~{\rm eV}$). Unfortunately, this is
impossible, because  it will cause too large corrections to
$\theta_{12}$. In Fig.~3, we depict the allowed regions in the plane
of $\tan\beta$ and the lightest neutrino mass $m_1$ in the NH case
(or $m_3$ in the IH case) by requiring $\theta_{12}(\Lambda_{\rm
EW})$ to be in its $1\sigma$, $2\sigma$ and $3\sigma$ ranges. As
shown in Fig.~3, both $\tan\beta$ and the lightest neutrino mass are
severely constrained. In the NH case, $\tan\beta$ can be relatively
large if $m_1 \lesssim 0.05~{\rm eV}$, whereas $\tan\beta < 20$ in
the IH case for the whole range of $m_3$. In the nearly degenerate
limit $m_1 \approx m_2 \approx m_3$,  we need a rather small
$\tan\beta$, which may lead to tensions with some supersymmetric
models.
\end{enumerate}

Finally, we stress that our discussions on the RGE's stick to the
effective theory approach featuring the tendency of generality. In a
class of seesaw models with flavor symmetries, the heavy seesaw
particles may possess non-degenerate masses. In this case, the RG
running between seesaw thresholds should be considered
accordingly~\cite{Antusch:2002rr,*Antusch:2005gp,*Bergstrom:2010id}.
In general, as long as one works in supersymmetric models, the
threshold effects should not be significant because of the
non-renormalization theorem. In the non-supersymmetric models, there
might be large threshold corrections to neutrino mixing angles due
to the existence of the Higgs self-coupling. However, the details of
the threshold behavior depend on the specific flavor structure of
the model, and in principle should be considered properly in model
buildings.

\section{Conclusions}

Recently, the long-baseline accelerator experiments T2K and MINOS
have observed $\nu_\mu \to \nu_e$ oscillations, indicating a
relatively large $\theta_{13}$, which will be soon tested in Double
CHOOZ and Daya Bay reactor neutrino experiments. The latest global
analysis of current oscillation data points to a best-fit value of
$\theta_{13} = 9^\circ$~\cite{Fogli:2011qn}. If such a
relatively-large $\theta_{13}$ is confirmed in the near future, our
understanding of the neutrino mixing pattern through flavor
symmetries would be dramatically changed. So far, various flavor
symmetries have been implemented to generate the tri-bimaximal
mixing pattern, which predicts $\theta_{12} \approx 35.3^\circ$,
$\theta_{23} = 45^\circ$ and $\theta_{13} = 0$. Now that a vanishing
$\theta_{13}$ is disfavored with a more than $3\sigma$ significance
\cite{Fogli:2011qn}, we are well motivated to consider significant
corrections to the tri-bimaximal mixing or to search for another
constant mixing pattern with a relatively large $\theta_{13}$.

In this paper, we concentrate on the so-called tetra-maximal mixing
pattern~\cite{Xing:2008ie}, which predicts $\theta_{13} \approx
8.4^\circ$ together with $\theta_{12} \approx 30.4^\circ$ and
$\theta_{23} = 45^\circ$. Although a relatively large $\theta_{13}$
is predicted, the solar mixing angle $\theta_{12}$ seems too small
to be consistent with current oscillation data. We propose two
feasible ways to solve this problem. First, since the tetra-maximal
mixing pattern can be written as a product of four maximal
rotations, one may introduce explicit perturbations to one or more
maximal rotation angles. We demonstrate that only one perturbation
parameter is enough to enhance $\theta^{} _{12}$ to its best-fit
value, and the maximal atmospheric mixing angle $\theta_{23} =
45^\circ$ and maximal CP-violating phase $\delta = -90^\circ$ are
maintained to any order of perturbations. Second, if the
tetra-maximal mixing pattern is produced by a certain flavor
symmetry at a high-energy scale $\Lambda = 10^{14}~{\rm GeV}$, the
radiative corrections governed by the RGE's can successfully enhance
$\theta_{12}$ to its best-fit value at the electroweak scale
$\Lambda_{\rm EW} = 10^2~{\rm GeV}$. We explicitly show that this is
really the case in the MSSM by solving the full set of RGE's. In
addition, $\theta_{13}$ and $\theta_{23}$, as well as three
CP-violating phases, are found to be rather stable against the
radiative corrections.

It is worthwhile to remark that the flavor symmetry underlying the
tetra-maximal mixing pattern deserves further studies. The
predictions for neutrino mixing parameters from the tetra-maximal
mixing pattern with or without corrections will be soon tested in a
number of precision neutrino oscillation experiments.

\vspace{-0.3cm}
\begin{acknowledgments}
The authors would like to thank Prof. Zhi-zhong Xing for
suggesting such an investigation and valuable discussions. This work was
supported by the ERC under the Starting Grant MANITOP and the
Deutsche Forschungsgemeinschaft in the Transregio 27 ``Neutrinos and
beyond -- weakly interacting particles in physics, astrophysics and
cosmology" (H.Z.) and by the Alexander von Humboldt Foundation
(S.Z.).
\end{acknowledgments}
\vspace{-0.3cm}


\ifx\mcitethebibliography\mciteundefinedmacro
\PackageError{apsrevM.bst}{mciteplus.sty has not been loaded} {This
bibstyle requires the use of the mciteplus package.}\fi

\end{document}